\documentstyle[12pt]{article}

\newcommand{\be}{\begin{equation}}
\newcommand{\ee}{\end{equation}}
\newcommand{\bn}{\begin{eqnarray}}
\newcommand{\en}{\end{eqnarray}}
\newcommand{\bd}{\begin{displaymath}}
\newcommand{\ed}{\end{displaymath}}

\begin{document}

\begin{flushright}
QMW-th/95-28
\end{flushright}

\begin{center}
{\Large \bf N=1 Heterotic-Supergravity Duality and\newline Joyce Manifolds}
\newline
\newline
\newline
{\large B.S. Acharya}\footnote{e-mail: acharya@qmw.ac.uk. Work Supported
By PPARC}
{\it Queen Mary and Westfield College,}
{\it Mile End Road,}
{\it London. E1 4NS}

\end{center}

\begin{abstract}

We construct the heterotic dual theory in four dimensions of
eleven dimensional supergravity compactified on a particular
Joyce manifold, $J$. In particular $J$ is constructed from resolving
fixed point singularities of orbifolds of the seven-torus in such
a way that one is forced to consider a generalised orbifold
compactification on the heterotic side. We conjecture that a
heterotic dual exists for all the compact 7-manifolds of
$G_2$ holonomy constructed by Joyce.
\end{abstract}
\newpage
\section{Introduction}
Our best candidates, to date, of fundamental descriptions of nature,
now all seem to be connected by the duality conjectures
\cite{duff,schwarz,hull,witten}. The emerging picture is one of an
underlying, even more fundamental, theory in which the other theories
emerge as one approaches various limits in the moduli space.

Much effort has gone into understanding the $N=4$ \cite{schwarz,hull},
$N=2$ \cite{sei,kach,klemm,louis} and more recently the $N=1$
\cite{seib,vaf,har} cases. In particular the $K3$ manifold plays a central role
\cite{hull,witten,ferr,strom,klemm,louis,vaf,paul,town,sen} in many
scenarios.

Compactification of $d=11$ supergravity on a
compact manifold of $G_{2}$ holonomy
gives an effective four dimensional $N=1$ theory. Joyce has given many
examples of such manifolds \cite{J1,J2}. It is natural to conjecture
that these manifolds play a crucial role in the $N=1$
truncations of the conjectured dualities
between the Type IIA superstring on $K3$, the heterotic string on $T^{4}$
and $d=11$ supergravity on $K3{\times}S^{1}$.

In \cite{har} a manifold of $G_{2}$ holonomy was constructed by considering
freely acting involutions of a $K3{\times}T^{3}$ orbifold and resolving the
singularities. These involutions translate directly to actions on
the heterotic side giving an example of $N=1$ duality.
However, freely acting involutions of this type are very limited, and
the case when the involutions act with fixed points on the seven
coordinates on the supergravity side is less well understood. In fact,
all of the manifolds of \cite{J1,J2} are constructed by resolving
fixed point singularities, and if these manifolds are to play
a role in $N=1$ duality then this situation needs to be understood.
This work aims to shed some light on this problem. We work at a
generic point in moduli space throughout the following.

We focus on the simplest example given in \cite{J1},
which we denote by $J$, constructed with
three non-freely acting involutions of the seven torus, the singularities
of which are then resolved. In section two a brief outline of Joyce's
construction is given. Then in section three we present the heterotic dual
compactification. In particular, we are forced
to consider a new kind of 'overlapping orbifold' on
the heterotic side of the duality map. We end with some conclusions.
\section{Joyce's Construction}

In \cite{J1,J2} Joyce gave explicit constructions of 7-manifolds of
$G_2$ holonomy. When eleven-dimensional supergravity is compactified
on these manifolds, the resulting theory is four dimensional N=1
supergravity with $b_2$ vector multiplets and $b_3$ chiral multiplets,
 where $b_2$ and $b_3$ are the non-trivial Betti numbers of the
 7-manifold. The examples of \cite{J1} are all constructed by
 orbifolding the seven-torus by various discrete isometries and then
 resolving the singularities by replacing them with non-compact
 Eguchi-Hanson geometries, a process that is now
 more than familiar to
 string theorists.

 The simplest example given in \cite{J1},
 which we denote by $J$ is constructed as follows:

 Define the seven-torus coordinates as $(x_1,......,x_7)$.
 Three $Z_2$ isometries of $T^7$ are defined by:
 \be
 \alpha(x_1,....x_7) = (-x_1,-x_2,-x_3,-x_4,x_5,x_6,x_7)
 \ee
 \be
\beta(x_1,....x_7) = (-x_1,1/2-x_2,x_3,x_4,-x_5,-x_6,x_7)
 \ee
 \be
 \gamma(x_1,....x_7) = (1/2-x_1,x_2,1/2-x_3,x_4,-x_5,x_6,-x_7)
 \ee

 Obviously each of these $Z_{2}$'s has 16 fixed singular $T^{3}$'s
 and each one defines an orbifold limit of a particular $K3{\times}
 T^{3}$.
There are thus 48 fixed singular $T^{3}$'s of the surface. However
one must ask how the other two isometries act on the singular set
of each $Z_{2}$. In particular,
the $Z_{2}{\times}Z_{2}$'s generated by
$(\beta,\gamma)$, $(\alpha,\gamma)
$ and
$(\alpha,\beta)$ act freely on the singular sets of $\alpha$,
 $\beta$ and $\gamma$ respectively. This implies that each isometry
has only four singularites of the original sixteen, giving a total
of 12 singular $ T^{3}$'s. Resolving each of these is crucial for the
consistency of the supergravity theory. This is done in the usual
way for $K3$ singularities - by inserting Eguchi-Hanson geometries, \cite{eh}.
The Betti numbers of the singular $T^{7}$ are $b_{2} = 0$ and $b_{3} = 7$.
The resolution of each singularity adds 1 to $b_{2}$ and 3 to $b_{3}$.
The Betti numbers of this Joyce manifold are thus $b_{2} = 12$
and $b_{3} = 43$.

\section{The Heterotic Dual and Generalised \newline Orbifold}

Eleven-dimensional Supergravity on $K3{\times}T^{3}$ is conjectured
to be dual to the heterotic string on $T^{6}$. Orbifolding the
$K3{\times}T^{3}$ by symmetries such as the Enriques involution
on the $K3$ part translates to a particular orbifold action
on the heterotic side. This has been illustrated successfully in
several examples \cite{ferr,sen,vaf,har}.

In particular, in \cite{har}
an example was considered in which $T^{7}$ was modded out
by a particular $Z_{2}{\times}Z_{2}{\times}Z_{2}$, one of
which defined a particular orbifold limit of $K3{\times}T^{3}$.
This is what has been called $\alpha$ in the construction of $J$.

The remaining $Z_{2}$'s acted freely, and we note that they are
essentially 'freely acting versions' of what we denoted by $\beta$
and $\gamma$ in constructing $J$. This is so because the example of \cite{har}
contained
half shifts of $S^{1}$'s defined by $x_{i} \rightarrow x_{i} + 1/2$.
In that example there is only one $K3$ present in constructing the
7-manifold and it is straightforward to map the freely acting
$Z_{2}{\times}Z_{2}$ to the heterotic side.
So what is the dual of the supergravity theory on $ J$?
$J$ was constructed by taking $T^{7}$, orbifolding by
three $Z_{2}$'s ({\sl each of which defines a particular $K3\times T^{3}$ })
and resolving all the singularities. In a sense, we have three
overlapping $K3$'s and when the singularities are suitably resolved
we give non-trivial holonomy to the whole $T^{7}$, promoting it to
$G_{2}$. This implies that on the heterotic side, the dual theory
should be 'an overlapping' of three $Z_{2}{\times}Z_{2}$ orbifolds,
because each of the three $K3$'s on the supergravity side should
be treated on an equal footing.

Denote by ${\alpha}\prime$, ${\beta}\prime$ and ${\gamma}\prime$ the
action of the $Z_{2}{\times}Z_{2}$
generated by $(\beta,\gamma)$, $(\alpha,\gamma)$ and $(\alpha,
\beta)$ respectively on the heterotic side. First we note that,
if treated separately, the $Z_{2}{\times}Z_{2}$ orbifolds given by
${\alpha}\prime$, ${\beta}\prime$ and ${\gamma}\prime$ each produce
the same massless spectra as the model condidered in \cite{har} \footnote
{Appropriate $S^{1}$ shifts are included in the
duality map, so that the adiabatic argument of \cite{vaf} applies
to each orbifold},
namely four vector multiplets and 19 chiral multiplets.

To find the massless spectrum of our 'overlapping orbifold', it is only
necessary to note that we have three copies of the same entity
on both sides of the duality map.{\sl On the supergravity side
resolution of singularities of {\it each} $K3$ gave a specific
massless spectrum. On the heterotic side, each of the three
orbifolds involved in the 'overlapping' each give essentially the same
compactification}. This suggests that
we sum the massless spectra separately,
which leads naturally to the definition
of the 'overlapping orbifold' as one in which
each orbifold involved in the process should
be treated separately, and the spectra summed.
However, we should only
count the dilaton and six moduli once. This gives a
spectrum with precisely 12 vector multiplets and 43 chiral
multiplets.
\section{Conclusions.}
It is further interesting to note that all the examples of Joyce
constructed from $Z_{2}$'s
have the Betti numbers of the singular $T^{7}$ as $b_{2} = 0$ and
$b_{3} = 7$. This should then always
correspond to the seven chiral multiplets
containing the dilaton and moduli. In our example, the number of singularities
of each isometry then corresponded on the heterotic side to
the number of surviving $ N=4$ vector multiplets, each of which
give rise to one $ N=1$ vector multiplet and three $ N=1$ chiral
multiplets on {\it both} sides.

Of the more general cases considered in \cite{J2}, it turns out
that for certain subsectors of the singularities there exists more
than one topologically distinct ways of resolving.
The different resolutions add different numbers to $b_{2}$ and $b_{3}$.
 This then should correspond on the heterotic side to subsets of $N=4$
 vector multiplets surviving the orbifold projection and possibly to
 extra massless states from the twisted sectors.
 It is thus natural to conjecture that
there is a heterotic dual for each of the manifolds given in \cite
{J2} and this deserves further investigation \cite{ba}.
\newline
{\bf Acknowledgements}.
The author is extremely indebted to Chris Hull,
Wafic Sabra, and Tomas Ortin for discussions and in particular
to Ashoke Sen for many
illuminating conversations.
The author would also like to thank PPARC, by whom this work is supported.

\end{document}